\documentclass[amsmath, amssymb, aip,jcp]{revtex4-1}
\usepackage[pdftex]{graphicx}
\usepackage{dcolumn}         
\usepackage{bm}              

\begin{document}

\title{A note on the metallization of compressed liquid hydrogen}

\author{Isaac Tamblyn} 
\email{itamblyn@dal.ca}
\affiliation{Department of Physics, Dalhousie University, Halifax, NS, B3H 3J5, Canada}
\author{Stanimir A. Bonev} 
\email{stanimir.bonev@dal.ca}
\affiliation{Department of Physics, Dalhousie University, Halifax, NS, B3H 3J5, Canada \\
                  Lawrence Livermore National Laboratory, Livermore CA 94550}
\date{\today}

\begin{abstract}

We examine the molecular-atomic transition in liquid hydrogen as it relates to metallization. Pair potentials are obtained from first principles molecular dynamics and compared with potentials derived from quadratic response. The results provide insight into the nature of covalent bonding under extreme conditions. Based on this analysis, we construct a schematic dissociation-metallization phase diagram and suggest experimental approaches that should significantly reduce the pressures necessary for the realization of the elusive metallic phase of hydrogen.

\end{abstract}


\maketitle


One of the fundamental questions relating to the properties of dense hydrogen is that of metallization. Since it was first predicted by Wigner and Huntington \cite{wigner_jcp_1935} in 1935, metallic hydrogen has long been sought after. Its value has only increased with time. The possibility of a high-T superconductor \cite{ashcroft_prl_1968,cudazzo_prl_2008}, or other ordered quantum states \cite{ashcroft_jpcm_2000}, has led to a tremendous amount of scientific investigation and thought. Despite this, many open questions remain as to the existence and properties of this exotic phase of matter.

Although it was originally estimated that metallization would occur at about 25 GPa, pressures ($P$) an order of magnitude larger have since been achieved in the lab and solid metallic hydrogen has yet to be observed. As new methods expanded the range of pressures attainable by experiment, theoretical predictions for the transition grew in step. Much has been written about the high pressures thought to be required to obtain metallic hydrogen \cite{mao_rmp_1994, johnson_nature_2000, stadele_prl_2000, loubeyre_nature_2002, nellis_rpp_2006, eremets_science_2008}. In a recent experimental study, Loubeyre \cite{loubeyre_nature_2002} and co-workers estimate that the transition pressure (for $T = $ 0 K) is in the vicinity of 450 GPa. Currently these conditions are outside of experimental reach. 

The notion that pressures on the order of several megabar are \emph{required} for metallization is potentially misleading, however. Indeed, metallic H has been experimentally realized already in the liquid phase at pressures as low as 140 GPa \cite{weir_prl_1996, fortov_prl_2007}. While this state was achieved for only a short time, it provides conclusive evidence for the existence of metallic state of H. Furthermore, it was recently proposed \cite{ashcroft_prl_2004}, and subsequently confirmed experimentally \cite{chen_pnas_2008}, that through ``chemical pre-compression'' hydrogen rich substances such as SiH$_4$ (silane) metallize at relatively low pressures (approximately 50-60 GPa). Promisingly, silane exhibits \cite{eremets_science_2008} a rapid increase in the superconducting transition temperature, $T_c$. These works highlight the fact that the conventional wisdom, essentially ``push harder'', may not be the best approach to obtaining metallic H.

In this article, we discuss the connection between molecular dissociation and metallization, a subject that has attracted considerable debate. Based on this consideration, we construct a schematic phase diagram for the liquid. We consider possible routes to the realization of the elusive phase of metallic H - ones that involve pressures lower than have been stated necessary, and at temperatures below what was required \cite{weir_prl_1996} experimentally.

While it is widely accepted that hydrogen (either as H or H$_2$) is an insulator at low densities, and should be metallic at extremely high densities, it is less clear where, and by what mechanism, the transition between these two phases occurs. At low densities, H atoms (or molecules) are sufficiently far apart that all possible atomic configurations are insulating. By insulating we mean that E$_{Fermi}$ falls below the Mott mobility edge \cite{mott_ap_1967}; at $T = $ 0 K the system will not conduct electricity. At extremely high densities, the opposite is true. \emph{All} possible structural configurations are metallic - electronic states at E$_{Fermi}$ are delocalized and conductivity is possible in the absence of electronic excitations. At intermediate densities, the situation is more complex.

\begin{figure}[!ht] 
\begin{center}
\includegraphics[width=0.9\textwidth,clip]{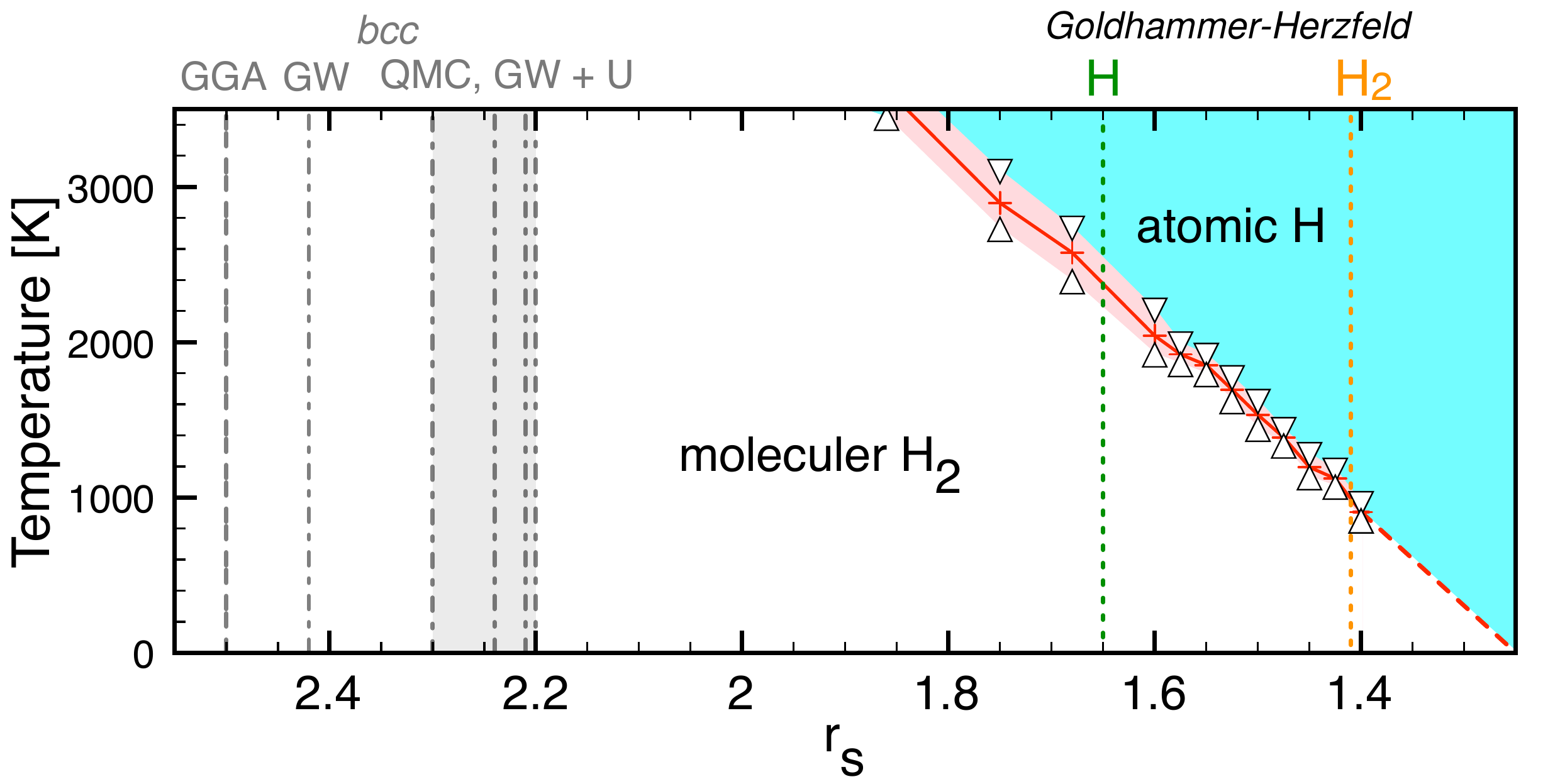}
\end{center}
\caption{\label{f:hydrogen_metallization_density} Predictions for the metallization density, $r_{metal}$, of hydrogen based on various theoretical approaches. For $r_s > r_{\mathrm{metal}}$, hydrogen will be an insulator. For $r_s < r_{\mathrm{metal}}$ it will be a metal. Predictions for $bcc$ atomic hydrogen are shown as grey vertical dashed lines \cite{zhu_thesis_1990, pfrommer_prb_1998, li_prb_2002}. An estimate based on the Goldhammer-Herzfeld criterion for atomic (green vertical dashed line) and molecular (orange vertical dashed line) phases are also indicated.
For comparison, a prediction for the molecular-atomic transition line \cite{tamblyn_prl_2009} based on the stability of molecular species in first principles molecular dynamics simulations is shown as a red solid line (dashed red line is an extrapolation). Upward and downward triangles bracket this transition based on molecular survival probabilities of 67\% and 33\% respectively (see \cite{tamblyn_prl_2009} for more discussion). For temperatures below the molecular-atomic transition line the liquid will be molecular, otherwise, it will be atomic.}
\end{figure}

To frame our discussion, we begin by reviewing results for a model system: $bcc$ atomic H. Depending on the approximations used, $bcc$ atomic H (which is not thermodynamically stable over the conditions discussed here, i.e. at least to $r_s$=1.30) is predicted to metallize at densities as low as $r_s = 2.78$ and as high as $r_s = 1.65$ \cite{zhu_thesis_1990, pfrommer_prb_1998, li_prb_2002}. Here the density is denoted in terms of the Wigner-Sietz radius, defined by $V/N =\frac{4}{3}\pi(r_{s}a_{0})^{3}$, where $V$ is the volume, $N$ the number of electrons, and $a_0$ is the Bohr radius. These values are plotted as grey vertical dashed lines in Fig.~\ref{f:hydrogen_metallization_density}. If we consider values only from the more exact methods (QMC and GW + U), $bcc$ atomic H is predicted to metallize within a range of densities corresponding to $r_s \approx 2.3 - 2.2$. Achieving such a density in the liquid or solid phase requires only a moderate degree of pressure ($P <$ 30 GPa, see for example \cite{desjarlais_prb_2003, scandolo_pnas_2003}).

As stated above, under these conditions an atomic $bcc$ phase is not thermodynamically preferred. Rather, this range of densities corresponds to phase \emph{I} hydrogen - a $hcp$ lattice of freely rotating H$_2$ molecules. This phase, along with phases \emph{II} and \emph{III}, is an insulating one. In fact, solid H has been shown to be an insulator at pressures up to 342 GPa by experiment \cite{narayana_nature_1998} and by theory alike \cite{stadele_prl_2000}. This is not a failing of the above predictions however, but rather illustrates a crucial point - metallization of hydrogen depends on whether the system is molecular or atomic.

In general, the larger the fraction of molecules present, the higher the metallization density will be. For liquid hydrogen, the degree of dissociation required for metallization is completely determined by the density of the system. One can define a parameter $r_{m_{X}}$ to indicate the density where a fraction $X$ of all atoms must be unpaired in order for the system to metallize. For $r_s > r_{m_{1.0}}$, the system will remain insulating even when all H$_2$ molecules are destroyed (by $T$ or some external means). Though subsequent heating may allow for conductivity by thermally excited carriers, the system will not be metallic. At most it will become a hot plasma. For $r_s < r_{m_{0.0}}$, the system (liquid or solid) will always be metallic. 

\begin{figure}[!tbh] 
\begin{center}
  \includegraphics[height=0.5\textwidth,clip]{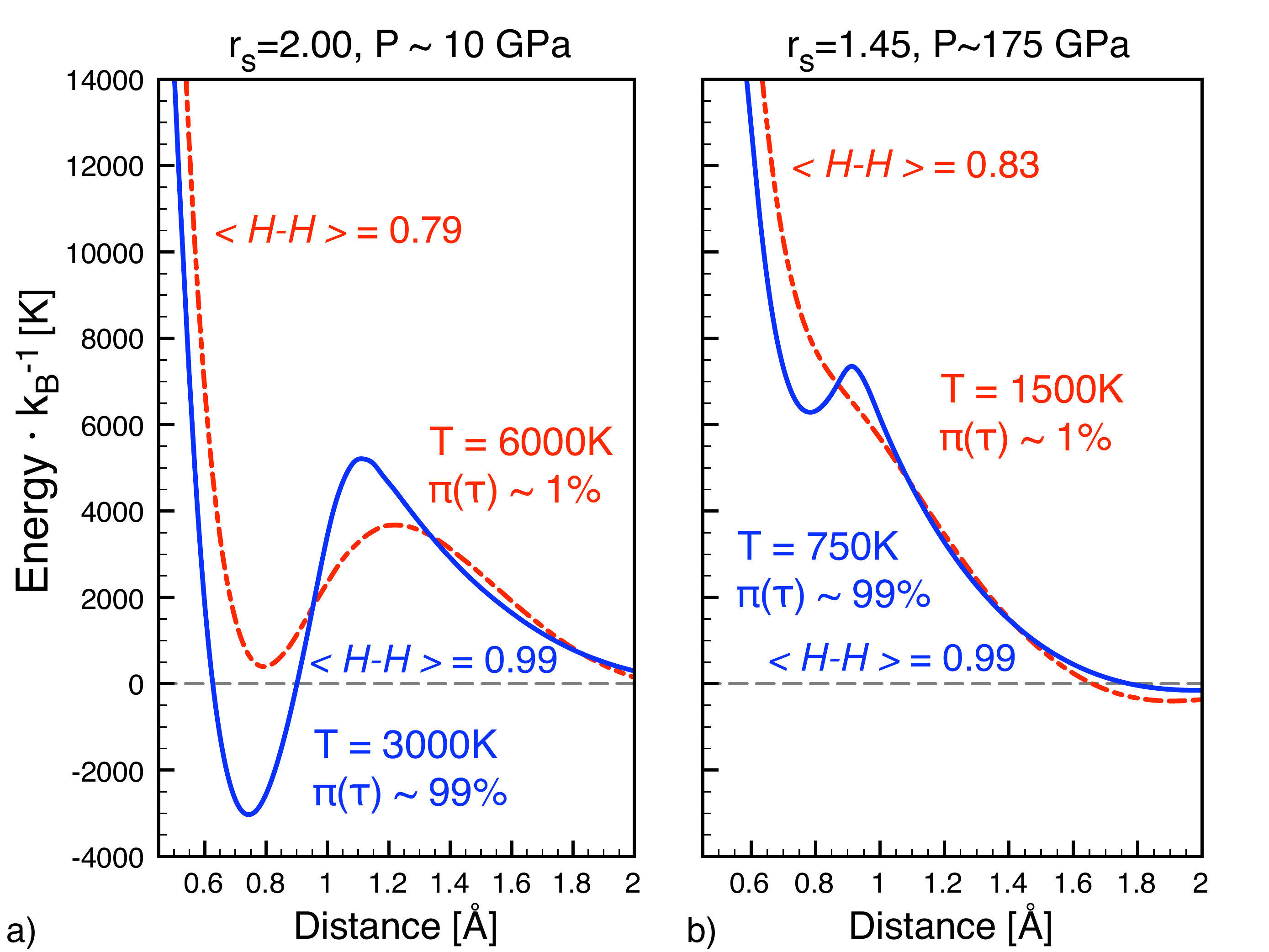}
\end{center}
\caption{\label{f:hydrogen_metallization_potentials} Effective two-body potentials derived (using the method of \cite{izvekov_jcp_2004}) from first principles molecular dynamics data. The survival probability of paired atoms, $\Pi(\tau)$, which relates directly to the degree of dissociation \cite{tamblyn_prl_2009}, is noted next to each curve. Additionally, we note the time average value of the fraction of mutual nearest neighbors, $\mathrm{<H-H>}$. This quantity describes the local structure of the liquid, but accounts poorly for changes in molecular stability. At lower densities, a), H-H interactions remain attractive in the dissociated (unstable) liquid, whereas at high density, b), they become repulsive. Small changes in the structure of the liquid (characterized by $\mathrm{<H-H>}$) can result in significant changes in interatomic interactions and hence stability when the liquid is highly compressed.}
\end{figure}

The relationship between metallization and dissociation can be probed using different theoretical approaches. Predictions of electrical conductivity \cite{holst_prl_2008, lin_prl_2009}, and polarizability \cite{tamblyn_prl_2009} are consistent with the picture of a metallic atomic fluid for pressures corresponding to $r_s < r_{m1.0}$. An additional approach to confirming metallization is to consider the nature of interatomic interactions that exist within the liquid. In Fig.~\ref{f:hydrogen_metallization_potentials}, we demonstrate the relationship between metallization and dissociation by constructing an effective interatomic potential using a force-matching technique \cite{izvekov_jcp_2004}. This approach uses configurations and forces obtained from first principles calculations to fit an averaged two-body potential. To generate relevant configurations, we performed large scale (as many as 1024 atoms) first principles molecular dynamics simulations in the vicinity of the molecular-atomic transition line over a wide range of densities. Simulations were carried out in the $N, V, T$ ensemble, where a Nose-Hoover thermostat was coupled to the ionic degrees of freedom and electronic states were populated according to the Fermi-Dirac distribution. The electronic density (within DFT-PBE) was optimized at each molecular dynamics step (i.e. Born Oppenheimer MD).

\begin{figure}[!h] 
\begin{center}
\includegraphics[height=0.5\textwidth,clip]{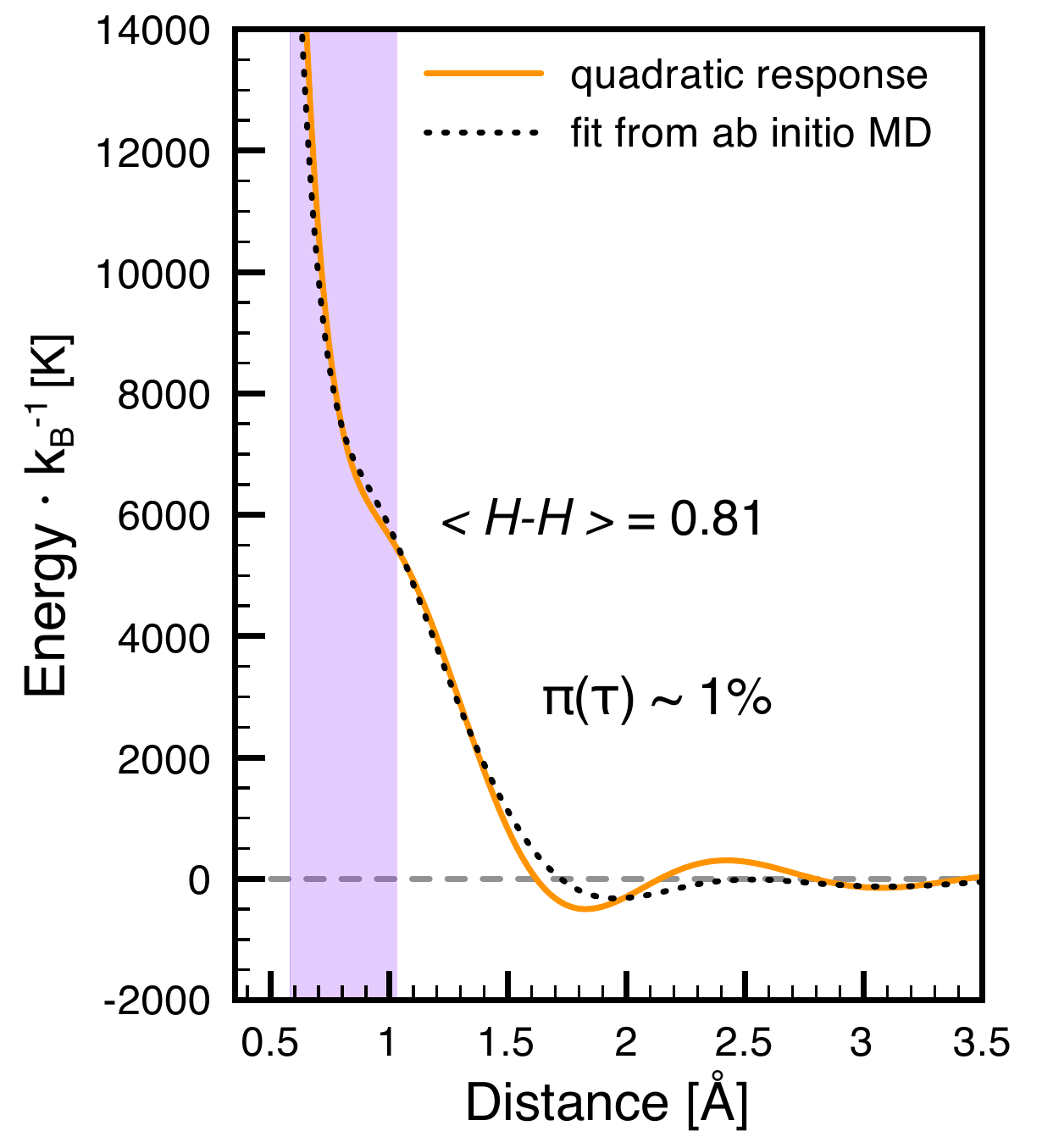}
\end{center}
\caption{\label{f:hydrogen_metallization_quadratic_response} Comparison of pair potentials generated from first principles molecular dynamics data (black dotted curve, $r_s$=1.50, $T$=2000K) and an analytic potential derived from quadratic response theory (orange solid curve). The analytic potential is independent of the structure of the liquid and depends only on the electronic density. The region corresponding to the 1$^{st}$ coordination shell is highlighted in purple.}
\end{figure}

From our effective potentials, we identify a change in interatomic interactions as the liquid is compressed. At low densities (Fig.~\ref{f:hydrogen_metallization_potentials}, a), H atoms are attracted to one another even when the system is highly dissociated (unstable). While the depth of the minima is somewhat reduced because of increased molecular scattering, it's overall character is essentially unchanged. This is consistent with an insulating atomic fluid. Conversely, for the high density case (Fig.~\ref{f:hydrogen_metallization_potentials}, b), thermal dissociation results in a marked change in interactions. The attractive region of the potential present at low $T$ completely disappears as the system is heated, even when only a vanishing fraction of molecules are destroyed (see caption of Fig.~\ref{f:hydrogen_metallization_potentials}). The resulting potential bears no resemblance to a molecular system.  H-H interactions are instead typical of a metallic system, where ion cores interact via a screened Coulomb repulsion. This fact can be seen more clearly by comparing our numerical potential with one derived using quadratic response (method described in \cite{nagao_prl_2003}), as in Fig.~\ref{f:hydrogen_metallization_quadratic_response}. This confirms not only the metallic character of the liquid, but suggests that such potentials, at least in the atomic regime, have little state dependence. We base this conclusion on the fact that our analytic potentials are derived without any structural information - the electronic density, which governs the degree of screening, is the only input variable. This implies that for the region of the phase diagram where the system is metallic, such potentials could reliably be used to accurately derive other physical properties. Furthermore, it would seem that they could be used with more exact methods such as Path Integral Monte Carlo, where both ions and electrons are treated as quantum mechanical particles. 

\begin{figure}[!h] 
\begin{center}
\includegraphics[width=0.9\textwidth,clip]{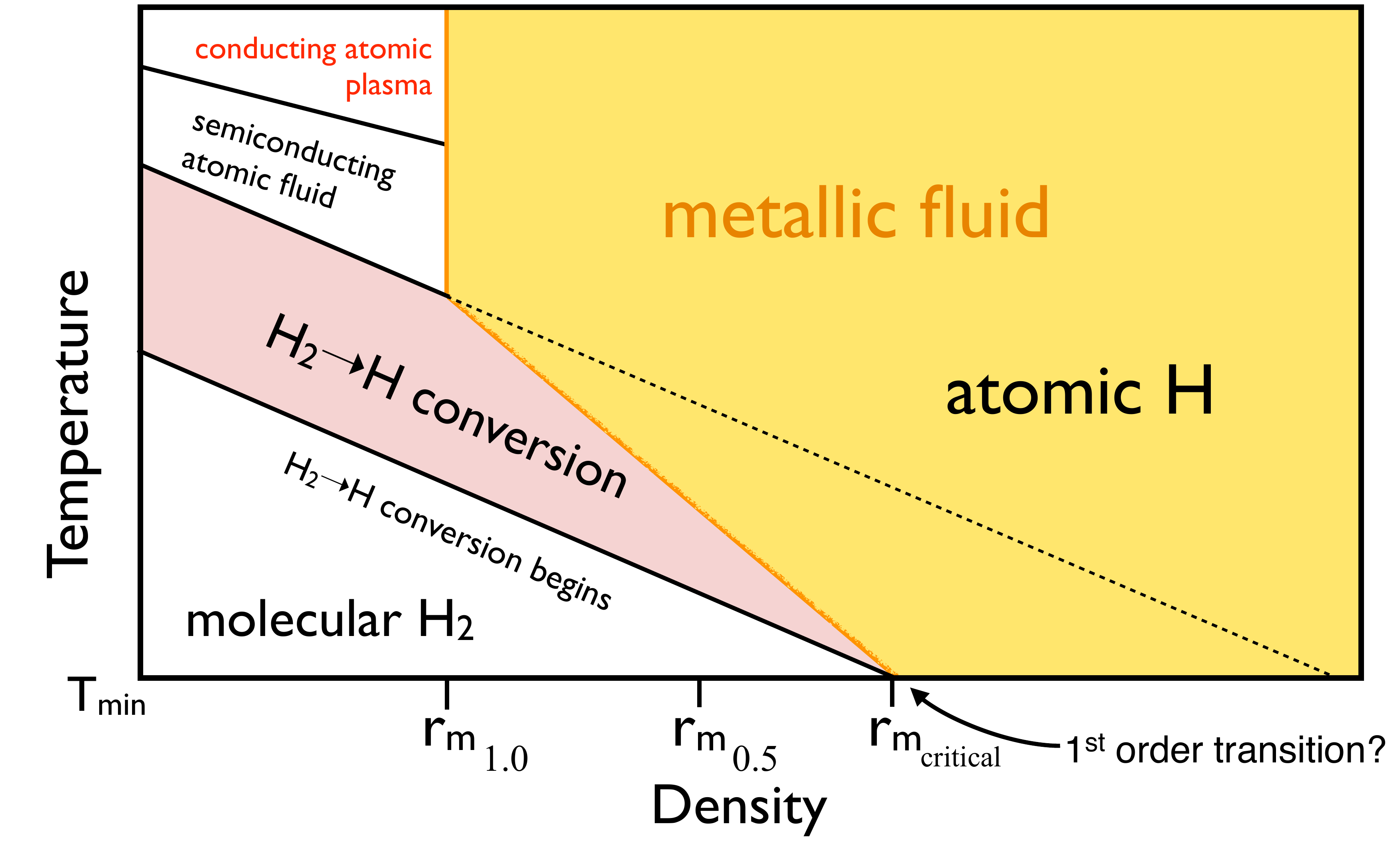}
\end{center}
\caption{\label{f:hydrogen_metallization_sketch} Sketch of the liquid hydrogen phase diagram over a wide range of conditions. The density corresponding to metallization for different degrees of dissociation is indicated $r_{m_X}$ (1.0 and 0.5 are shown). As density increases, the degree of dissociation required for metallization decreases. Yellow shaded region indicates where the liquid should be metallic. Insulating-metallic transition is indicated by an orange line. At fixed density the liquid can sample both insulating and metallic configurations, therefore this transition is continuous. Black dashed line indicates the molecular-atomic phase boundary that would exist if metallization did not occur. $T_{min}$ is the minimum temperature at which the liquid is stable (either $0 = K$ or the melting temperature of the solid phase). The intersection between the H$_2\rightarrow$ 2H line and the orange metallization line is point where the onset of dissociation and metallization would be coincident. This should result in a discontinuous molecular-atomic transition. We note that other first-order transitions (\emph{e.g.} structural \cite{tamblyn_prl_2009}) may also occur at lower densities.}
\end{figure}

Based on these considerations, we can now construct a dissociation-metallization schematic phase diagram (Fig.~\ref{f:hydrogen_metallization_sketch}). In the limit of $P \approx 0$, molecular dissociation is governed by the competition of the energy cost of dissociation and the corresponding entropic gain. In this regime, the system undergoes a series of thermally induced transitions, from an insulating molecular fluid, to a semiconducting atomic fluid, until finally it becomes a conducting atomic plasma. In this case, conversion of H$_2\rightarrow$ 2H occurs continuously. 

As hydrogen undergoes compression from vanishing densities, the temperature required to dissociate molecules initially increases. Upon sufficient compression, however, many-body interactions will lower the dissociation energy and thus the temperature required for H$_2 \rightarrow $ 2H conversion. In condensed systems, the slope of the molecular-atomic transition line is therefore negative.

At densities where $r_s < r_{m_{1.0}}$ this transition (H$_2 \rightarrow$ 2H) is affected by metallization. Metallization results in charge delocalization which in turn inhibits intramolecular bonding through screening of the ion cores. This has the effect of destabilizing H$_2$ bonds. For a system at a density of $r_{m_X}$, once X of the molecules have been thermally destroyed, metallization will occur and all remaining H$_2$ will dissociate. 

Since the dissociation fraction necessary for metallization decreases with density, at some point the molecular-atomic and the insulator-metallic transition lines must merge. This is the point where only an infinitesimal fraction of H$_2$ need dissociate for the system to metallize. This transition should be abrupt, and may be discontinuous, \emph{i.e.} 1$^{st}$ order. We note that that the occurrence of a such a transition does not preclude the possibility of other structural transitions at lower pressures \cite{tamblyn_prl_2009}.  Beyond $ r_{m_{0.0}}$, the liquid is expected to be metallic for all $T$. A sketch indicating the location of this transition, along with all other phase boundaries we have discussed is given in Fig.~\ref{f:hydrogen_metallization_sketch}.

The fact that dissociation is not necessarily equivalent to metallization may partially explain the large difference in temperature between predictions of the molecular-atomic transition and measurements of the semiconducting-metal transition \cite{weir_prl_1996, nellis_science_1996, nellis_prb_1999}. Indeed, Hood and Galli \cite{hood_jcp_2004} showed that for the case of shockwave experiments, metallization occurs when the fraction of H$_2$ present is approximately 0.60, not 0.05 as was initially proposed (see note \cite{hood_note} and Ref.~\cite{marvin_prb_1996}).

We return now to the issue of pressure and metallization. Pressurization serves several distinct purposes. At ambient conditions, the density of hydrogen is below the critical density necessary for metallization (i.e. $r_s > r_{\mathrm{metal}}$). Application of moderate pressure is therefore necessary to compress the system so that  $r_s < r_{\mathrm{metal}}$. Once this density has been achieved, metallization can be induced by two distinct mechanisms: increasing the fraction of atomic H through dissociation or increasing the density of the system, thereby lowering the atomic fraction necessary for metallization. An increase in density also has the secondary effect of reducing the dissociation energy of H$_2$; covalent bonding is inhibited by screening that results from many-body interactions. These mechanisms may act in concert, as for example in a shockwave experiment \cite{nellis_science_1996}. When a shockwave propagates through a sample, matter it encounters is compressed and heated. Compression reduces the atomic fraction necessary for metallization, while heating promotes dissociation due to the increase in molecular vibrational energy.

Other possible routes to dissociation, and therefore metallization, include experiments that induce electronic or vibrational transitions (e.g. an optical probe). Work by Nagao \emph{et. al} \cite{nagao_prl_2003} suggests that in the solid phase, development of intermediate range crystalline order extends the range of pressure over which molecular phases are thermodynamically preferred. The corollary of this hypothesis is that at high densities, disruption of this order by way of chemical doping should aid in the dissociation of molecules. Similarly, metastable phases obtained by quenching the liquid may be more conducive to metallization. Promotion of collective excitations such as plasmons will also reduce the stability of molecular species. Injection of hot electrons could also be used to promote molecular destabilization. Subjecting the system to strong electric fields or forcing charge transfer \cite{zurek_pnas_2009, tamblyn_thesis_2009} may also be a viable approaches. This could be achieved by donation from electropositive species \cite{zurek_pnas_2009} or by photonic emission from metal surfaces with low work functions.

By the same token, it should be possible to suppress metallization by inserting neutral additives (e.g. helium) into the liquid, effectively lowering the concentration of the hydrogen subsystem. Finally, we note that additives which catalyze H$_2 \rightarrow $ 2H should aid in metallization, provided this effect is larger than the corresponding dilution. For example, transition metal atoms have been proposed as a means of dissociating H$_2$ \cite{carlsson_prl_1983}. In such cases it will be crucial to maintain a well defined hydrogen subsystem - metallic hydrogen is not simply hydrogen in the presence of a metal. 

More recently, experimental work on silane/H$_2$ mixtures \cite{strobel_prl_2009} indicates that under pressure, SiH$_4$ has the effect of weakening H-H bonds. At the highest pressures observed, the H-H vibron was found to be 4090 cm$^{-1}$. As the authors point out, to soften the H-H stretch mode to this extent in pure H$_2$ requires pressures nearly 4 times as large. This result is encouraging, and suggests that metallic H may finally be within reach.

Any combination of the processes we have discussed (as well as ``chemical pre-compression'') should further reduce densities, and therefore pressures, required to achieve metallization. We therefore suggest that future high pressure experimental work focus on techniques aimed at promoting dissociation of H$_2$, as it will ultimately lead to metallization. 
\section{Acknowledgments}
Work supported by NSERC and CFI. Compute resources were provided by ACEnet. I.T. acknowledges support by the Killam Trusts. Work at the Lawrence Livermore National Laboratory was performed under Contract DE-AC52-07NA27344.
%

\begin{thebibliography}{10}
\bibitem{hood_note}%
  \BibitemOpen
  \bibinfo {note} {The degree of dissociation quoted in this work is based on
  the location of Wannier centers relative to H atoms. This is definition is
  different than the one used in Fig \ref{f:hydrogen_metallization_density},
  i.e. molecular stability.}\BibitemShut{Stop}%
\end{thebibliography}
%

%
\end{document}